\documentclass[12pt,preprint]{aastex}
\shorttitle{Secondary Masses in PCEBs}
\newcommand{\Msun}{M_\odot}
\newcommand{\Rsun}{R_\odot}

\begin{document}

\title{The Distribution of Secondary Masses in Post-Common Envelope Binaries:  A Potential Test of Disrupted Magnetic Braking}

\author{Michael Politano and Kevin P. Weiler}

\affil{Department of Physics, Marquette University, P.O. Box 1881, Milwaukee, WI 53201-1881}

\begin{abstract}
The distribution of secondary star masses in present-day post-common envelope binaries (PCEBs) is calculated using four different models for angular momentum loss (AML) during the post-CE phase: only gravitational radiation (GR), GR + disrupted magnetic braking (DMB), GR + reduced MB, and GR + intermediate MB.  For the DMB model, we find that the number of PCEBs decreases abruptly by 38\% once MB begins to operate for non-fully convective secondaries.  We do not find a similar feature in the distributions calculated using any of the other three AML models in which MB is not disrupted.  This percentage decrease in the number of present-day PCEBs predicted using the DMB model is easily large enough so that an observed distribution of secondary masses or even spectral types in PCEBs can provide an important test of whether magnetic braking is indeed substantially reduced in secondary stars that are fully convective.  We discuss briefly the feasibility of such observations.  
\end{abstract}


\section{Introduction}

The angular momentum loss (AML) mechanisms that operate to drive mass transfer in cataclysmic variables (CVs) have been the subject of debate recently.  Historically, it has been believed that gravitational radiation (GR) is the main driver for CVs having orbital periods of 3 hours or less and magnetic braking (MB) the main driver for orbital periods of 3 hours or greater.  In the orbital period distribution of observed CVs, there is a significant shortage of systems with periods between 2 and 3 hours.  This feature is referred to as the ``period gap."  A generally accepted model for the period gap, proposed independently by Rappaport, Verbunt and Joss (1983; hereafter referred to as RVJ) and \citet{spr83}, is that MB is severely reduced once the secondary star becomes fully convective.  This occurs for a mass of $\sim$ 0.3-0.4 $\Msun$, which corresponds to an orbital period in CVs of $\sim$ 3 hours.  Once MB ceases, the secondary, whose radius is slightly larger than its main sequence radius due to the high mass transfer rates caused by MB, relaxes back to its thermal equilibrium radius and becomes detached.  Mass transfer recommences once GR brings the Roche lobe back into contact with the secondary at $\sim$ 2 hours.  This model has become known as the disrupted magnetic braking (DMB) model.

Recently, the DMB model has been called into question by observations of magnetic activity in single stars in young open clusters \citep{sil00,pin02}.  \citet{sil00} find no significant decrease in magnetic activity through the mass range where the secondary becomes fully convective.  They also find that the magnetic activity saturates at stellar rotation rates larger than some critical rotation rate, $\Omega_{crit}$, and claim that the prescription used in RVJ overestimates the strength of the AML due to MB at high rotation rates.  \citet{and03} have constructed a MB prescription that better fits the observational data from open clusters.  In their model, $\dot{J}_{MB}$ varies as $\Omega^3$ (as in RVJ) for rotation rates slower than $\Omega_{crit}$, and as $\Omega$ for faster rotation rates.  This model, which has become known as ``reduced magnetic braking (RMB),'' if applicable to the secondary stars in CVs, poses serious problems for the DMB model.  However, we should also note that the RMB model provides no alternate explanation of the period gap or of the observed brightnesses of CVs above the period gap.  

\citet{iva03} have recently proposed a MB prescription based on a two-component coronal model of stellar magnetic fields \citep{mes87} and on observations of X-ray emission from rotating dwarfs that provide an empirical relationship between stellar activity and rotation \citep{piz03}.  Their model is similar to the RMB model in that there is a different dependence upon $\Omega$ for fast rotators than for slow rotators.  However, the dependence of $\dot{J}$ on $\Omega$ in their model for fast rotators is a bit stronger than in RMB (see section 2.2.).   Hence, we shall designate this model as ``intermediate magnetic braking (IMB).'' 

The same AML mechanisms that drive mass transfer in CVs also act prior to the CV phase, when the binary is detached.  Close binary stars in this stage of their evolution are typically referred to as ``post-common envelope binaries (PCEBs).'' To avoid any potential ambiguity, we note that in this paper the term PCEBs specifically refers to binaries that (1) are detached, (2) contain a white dwarf (WD) primary and a main-sequence or brown dwarf secondary, and (3) have undergone a single CE phase.  We present in this paper population synthesis calculations of the secondary mass distribution in present-day PCEBs for the DMB, RMB and IMB models, as well as for no MB (i.e., only GR).  We show that this distribution can provide a relatively straightforward observational test of whether magnetic braking is indeed disrupted once the secondary star becomes fully convective.  

We note that most early population synthesis calculations of PCEBs focused on exploring statistically the possibility that newly-formed PCEBs could be responsible for bipolar planetary nebulae \citep{dek90,dek93,yun93,han95}.  With one exception, these studies did not model the subsequent evolution of the newly-formed PCEBs and thus did not consider AML mechanisms.  Only \citet{dek93} calculated models of the present-day PCEB population using DMB and we compare our results with theirs in section 4.  Recent investigations of PCEBs have been done by \citet{sch03}, \citet{han03}, \citet{wil04}, and \citet{nel05}, but none are suitable for comparison.  \citet{sch03} did not perform full population synthesis calculations, \citet{wil04} computed formation models of PCEBs, but not present-day, evolved models including an AML phase, \citet{han03} only considered sdB binaries, and \citet{nel05} were mainly concerned with  reconstructing the evolution of observed PCEBs through prior phases of mass transfer and assumed negligible orbital evolution of the PCEBs due to AML.

\section{Method}

\subsection{Population Synthesis Code}

The Monte Carlo population synthesis code used in this study is the same as the one used in \citet{pol04} to model ZACVs.  This code is described in \citet{pol88,pol96,pol04}, and we must refer the reader to those papers for detailed discussions.  Here we summarize only the key assumptions and features of the code and discuss how the code was used to calculate the present-day PCEB population. 

For all calculations, we begin with 10$^7$ zero-age main sequence (ZAMS) binaries.  Following a standard approach, we assume that the distribution of these ZAMS binaries can be written as a product of three separate distributions over primary mass, mass ratio, and orbital period.  We use a \citet{mil79} distribution for the primary masses, a distribution that is flat in q for the mass ratios (i.e., g(q) dq = 1 dq, where q = M$_s$/M$_p$, \citealp{duq91,maz92,gol03}), and a distribution that is flat in log P for the orbital periods \citep{abt83}.  For a given primary mass, the secondary mass is chosen from the distribution, F(M$_s$) = f(M$_p$)$\,$g(q), where f(M$_p$) is the above-mentioned \citet{mil79} distribution used for the primary masses.  

To model the population of PCEBs, a given ZAMS binary is evolved to the point where the primary contacts its Roche lobe during its ascension of one of the giant branches.  Wind loss during ascent of the giant branch(es) is incorporated via a \citet{rei75} prescription.  Relationships between the radius of the giant, its core mass and its total mass are derived from analytic fits to detailed stellar evolution models \citep{pol88,pol96}.   Simple energetic considerations are used to relate the pre- and post-CE orbital separations for the CE phase and a standard constant $\alpha_{CE}$ prescription is used with $\alpha_{CE}$ = 1 (see \citealp{tut79,pol04}).  The classical prescription for GR \citep{lan51} and various prescriptions for MB (see next section) are used to describe AML during the post-CE phase.  We use detailed stellar models from the Lyon group \citep{cb97,bar98,bar03,cha00} for low-mass secondaries ($\la$ 0.5 $\Msun$) and fits to stellar models from Webbink (see \citealp{pol88,pol96}) for secondaries with masses greater than this.  In our models, the secondary is fully convective for M$_s\leq\,$0.37 $\Msun$.  To avoid any possible confusion with CV evolution, we emphasize that the secondary stars in PCEBs do not lose mass as a result of mass transfer since the binary is detached.  Therefore, the secondaries in PCEBs do not ``become'' fully convective.  They either are fully convective or are not fully convective in our models, depending on whether their mass is less than 0.37 $\Msun$ or greater than 0.37 $\Msun$, respectively.  Finally, we terminate MB in all AML models once the secondary develops a radiative envelope at 1.25 $\Msun$.
 
In calculating the population of present-day PCEBs, four timescales are of interest: (1) $t_b$, the time that the progenitor binary was formed (measured from the beginning of the Galaxy), (2) $t_{ev,p}$, the time it takes the primary to evolve off of the main sequence, become a giant, and contact its Roche lobe to initiate the CE phase, (3) $t_{PCEB}$, the time from the end of the CE phase until the present epoch, and (4), $t_{Gal}$, the age of the Galaxy.  These four timescales must satisfy the following constraint,
$t_{b} + t_{ev,p} + t_{PCEB} =  t_{Gal}$.
We make the following assumptions regarding these time scales: (1) the stellar birth rate throughout the Galaxy's history has been constant, (2) the CE phase is so rapid that the time spent in it by the binary is negligible compared to the other time scales \citep{mey79}, and (3) the age of the Galaxy is 10$^{10}$ yrs.

\subsection{AML Prescriptions}

We use four different AML prescriptions that differ in how MB is modeled.  Model 1 is no MB, only GR.  We use the standard prescription for GR \citep{lan51}:
\begin{equation}
\dot{J}_{GR} =  -\frac{32}{5}\, \frac{G}{c^5}\,
\left( \frac{M_1M_2}{M_1 + M_2} \right) ^2\, a^4\, \Omega^5 ,
\end{equation}
where G is the gravitational constant, c is the speed of light, a is the orbital separation, M$_1$ and M$_2$ are the masses of the two stars, and $\Omega$ is the orbital angular frequency.  Model 2 is GR + RMB, where the prescription for RMB is given by \citet{and03}:
\begin{equation}
\dot{J}_{RMB} =  -7.3\times10^{30}\,dyn\,cm \left(\frac{R}{\Rsun}\right)^{0.5} 
\left(\frac{M}{\Msun}\right)^{-0.5} 
\cases{(\Omega/\Omega_\odot)^3, & if $\Omega \leq \Omega_{crit}$; \cr 
(\Omega_{crit}^2\Omega)/\Omega_\odot^3, & if $\Omega > \Omega_{crit}$, \cr}
\end{equation}
and where M and R are the mass and radius, respectively, of the secondary star and $\Omega_{crit}$ is the rotation rate at which MB saturates.  Model 3 is GR + IMB, where the prescription for IMB is given by \citep{iva03}:
\begin{equation}
\dot{J}_{IMB} =  -6\times10^{30}\,dyn\,cm \left(\frac{R}{\Rsun}\right)^{4} \left(\frac{T_d}{T_{d,\odot}}\right)^{0.5} \cases{(\Omega/\Omega_\odot)^3, & if  $\Omega \leq \Omega_x$; \cr 
(\Omega_x^{1.7}\,\Omega^{1.3})/\Omega_\odot^3, & if $\Omega > \Omega_x$. \cr}
\end{equation}
In this prescription, $\Omega_x$ is the rotation rate at which the dependence of the X-ray luminosity on $\Omega$ changes form, which we assume is equal to 10 $\Omega_\odot$, and we assume T$_d$ is equal to T$_{d,\odot}$ (\citealp{iva03}; R. Taam, priv. comm.).
Finally, model 4 is GR + DMB, where the prescription for DMB given by RVJ is used:
\begin{equation}
\dot{J}_{DMB} =  \cases{-4.8\times10^{30}\,dyn\,cm \;\,(R/\Rsun)^2
(M/\Msun) (\Omega/\Omega_\odot)^3, & if  $M > 0.37\,\Msun$; \cr
0, & if $M \leq 0.37\,\Msun$. \cr}
\end{equation}

\section{Results}

In Figure 1, we show the distribution of secondary masses in present-day PCEBs for each of the four AML models described in section 2.2. The dotted line is the distribution for GR only, the dashed line for GR + RMB, the dashed-dotted line for GR + IMB, and the solid line for GR + DMB.  The scale on the y-axis has been arbitrarily normalized to facilitate the discussion below.  

The most striking feature in this figure is the sharp drop in the number of present-day PCEBs once MB is turned on at M$_s$ = 0.37 $\Msun$ in the DMB model.  The distributions calculated using the other AML prescriptions do not show a similar feature.  This sharp drop is due to the increased efficiency of MB to bring systems into contact compared to GR in the DMB model.   Rewriting equation 1 in units similar to equation 4, we have
\begin{equation}
\dot{J}_{GR} =  -6.2\times10^{29}\,dyn\,cm\;
\left(\frac{M_1M_2}{[M_1 + M_2]^{1/3}}\right)^2 
\left(\frac{\Omega}{\Omega_\odot}\right)^{7/3}.
\end{equation}
Comparison of equations 4 and 5 shows that $\dot{J}$ is $\sim$ 50 times greater for MB than for GR in the DMB model (assuming a typical WD mass, M$_1$ = 0.6 $\Msun$).   As a result, when MB is operating, PCEBs are brought into contact relatively quickly following the CE phase and become CVs before the present epoch is reached.  On the other hand, if only GR is operating, many more PCEBs remain detached at the present epoch. 

The percentage decrease in the number of PCEBs once MB is turned on in the DMB model is 38\%.  This percentage decrease compares quite favorably with that found by \citet{dek93} ($\sim\,$44\%, see their Fig. 4c), who also calculated the secondary mass distribution in present-day PCEBs using DMB.   

For the DMB model, we also investigated the effects of different values of $\alpha_{CE}$ and different ZAMS mass ratio distributions on the resulting present-day PCEB secondary mass distribution.  In addition to $\alpha_{CE}$ = 1 and g(q) = 1, we calculated distributions for $\alpha_{CE}$ = 0.6, 0.3 and 0.05, and for g(q) $\propto$ q and g(q) $\propto$ q$^{-0.9}$.  We find that while the total number of PCEBs is sensitive to the choice of $\alpha_{CE}$, the percentage decrease in the number of PCEBs at 0.37 $\Msun$ remains relatively unaffected for $\alpha_{CE}$ between 1.0 and 0.6, and then increases from 37\% to 73\% as $\alpha_{CE}$ is reduced to 0.05.  We also find that while the overall shape of the distribution is affected by the choice of g(q), the sharp drop at M$_s$ = 0.37 $\Msun$ remains prominent in each distribution, with only a slight variation in the percentage decrease of PCEBs (29\% to 37\%).

\section{Discussion}

In our population synthesis calculations using the DMB model, MB does not operate for secondary stars that are fully convective (M$_s$$\,\leq\,$0.37 $\Msun$ in our models).  For secondary stars that are not fully convective (0.37$\,$$\Msun$ $<$ M$_s$ $<$ 1.25$\,$$\Msun$ in our models), MB does operate in the DMB model.  The number of present-day PCEBs predicted using the DMB model decreases abruptly by 38\% when magnetic braking is turned on at M$_s$ = 0.37 $\Msun$.  The present-day PCEB secondary mass distributions predicted using the other three AML models, in which MB is either always off (GR only) or always on (RMB and IMB) show no similar abrupt decrease.  Notably, the percentage decrease predicted using the DMB model is insensitive to the assumed choices for the common envelope efficiency parameter and the distribution of mass ratios in ZAMS binaries, two input factors whose uncertainties have notoriously plagued population synthesis calculations of close binary systems.  Such insensitivity implies that this prediction is fairly robust.  Consequently, observations of PCEBs with secondary masses in the range $\sim$ 0.25 - 0.5 $\Msun$ (approximately  M and late K spectral types) should provide an important test of whether MB is indeed disrupted once the secondary becomes fully convective.  

As few as six years ago, there were only $\sim$ 40 known PCEBs \citep{hil00} and the prospects for observational comparison with theoretical models might have seemed bleak.  However, with the advent of large observational surveys such as the Sloan Digital Sky Survey (SDSS), the observed number of detached WD/MS binaries has increased substantially.  \citet{sil06} have recently extended the work of \citet{ray03} and have compiled a catalog of 746 spectroscopically-identified, detached close binary systems containing a WD and a low-mass MS secondary star in the SDSS through the Fourth Data Release \citep{ade06}.  They find that the vast majority of these systems ($\sim$ 700) contain M dwarf secondaries and that the distribution of secondary spectral types in these binaries peaks at a spectral type of M4 and trails off quickly for earlier and later spectral types (see Fig. 10a in \citealp{sil06}).  However, the majority ($\sim$ 80\%) of the detached WD/MS binaries in the \citet{sil06} sample were targeted for spectroscopy because their colors resembled other, higher priority objects, such as quasars.  \citet{ric02} discuss in detail how quasar candidates were selected in the SDSS and the regions in color space that were excluded because of potential contamination of the quasar sample.  One such excluded region is precisely where WD-M dwarf binaries reside.  Consequently, as noted in \citet{sil06}, their sample is neither well-defined photometrically nor statistically complete and should not be considered as representative of the secondary mass function in PCEBs. 

Nevertheless, this recent large increase in the number of detached WD/MS binaries is encouraging and suggests that the test of the DMB model we propose is feasible obervationally.  This feasibility is greatly aided by the following factors: (1) the predicted percentage decrease in the number of PCEBs in the DMB model is large enough so that a modest sample ($\sim$ 50-100 PCEBs) should suffice; (2) the range of secondary masses to be targeted is narrow and fairly well-defined ($\sim$ 0.25-0.50 $\Msun$); (3) precise determinations of the secondary masses are not necessary, spectral types will suffice, and (4) precise orbital periods are not needed, only a coarse determination of whether the system could have undergone a CE phase (i.e., has an orbital period less than $\sim$ 10 days).  

An observational challenge that will need to be addressed is the detection of PCEBs containing either a hot WD and a very low mass secondary or a cool WD and an early M/late K secondary.  In the former case, the WD is much brighter than the secondary in the optical and the system may be mistaken for a single WD based on its colors.  In the latter case, the WD is hidden by the glare of the secondary, and the system may be misidentifed based on its colors as a single early M or K star.  For example, \citet{sch03} computed the colors expected for PCEBs containing a WD in the temperature range 6000 - 35,000 K and a ZAMS secondary with spectral type K0 to M6.  They compared their simulated colors with the U - B selection criteria for the Palomar-Green survey \citep{gre86} and concluded that PCEBs containing a relatively cool (T$_{eff}\,<\,$15,000 K) WD would have been included in the Palomar-Green survey only if they had a companion of spectral type $\sim$ M4 or later.  We note that observations in the UV or blue region of the optical spectrum ($\sim$ 3500-5000 $\mathaccent'27 A$), where early M dwarfs contribute very little flux, may help to reveal a WD companion if it exists.  However, optical observations in the blue would need to be at a higher resolution than SDSS to achieve an acceptable signal to noise ratio (N. Silvestri, priv. comm.).  EUVE or X-ray observations may prove useful in revealing K stars that are hiding a WD companion (e.g., \citealp{gre00,goo05}).

\acknowledgements {Warm thanks to Drs. C. Carden, D. Drapes, D. Hoard, N. Silvestri, K. Simkunas, P. Szkody, R. Taam and B. Willems for very useful discussions about this work.  We are particularly grateful to Dr. N. Silvestri for providing access to her SDSS observations in advance of publication.  Finally, we thank the referee for very helpful comments that improved the paper.  This work was funded in part by NSF grant AST-0328484 to Marquette University.}

\clearpage


\begin{figure}
\plotone{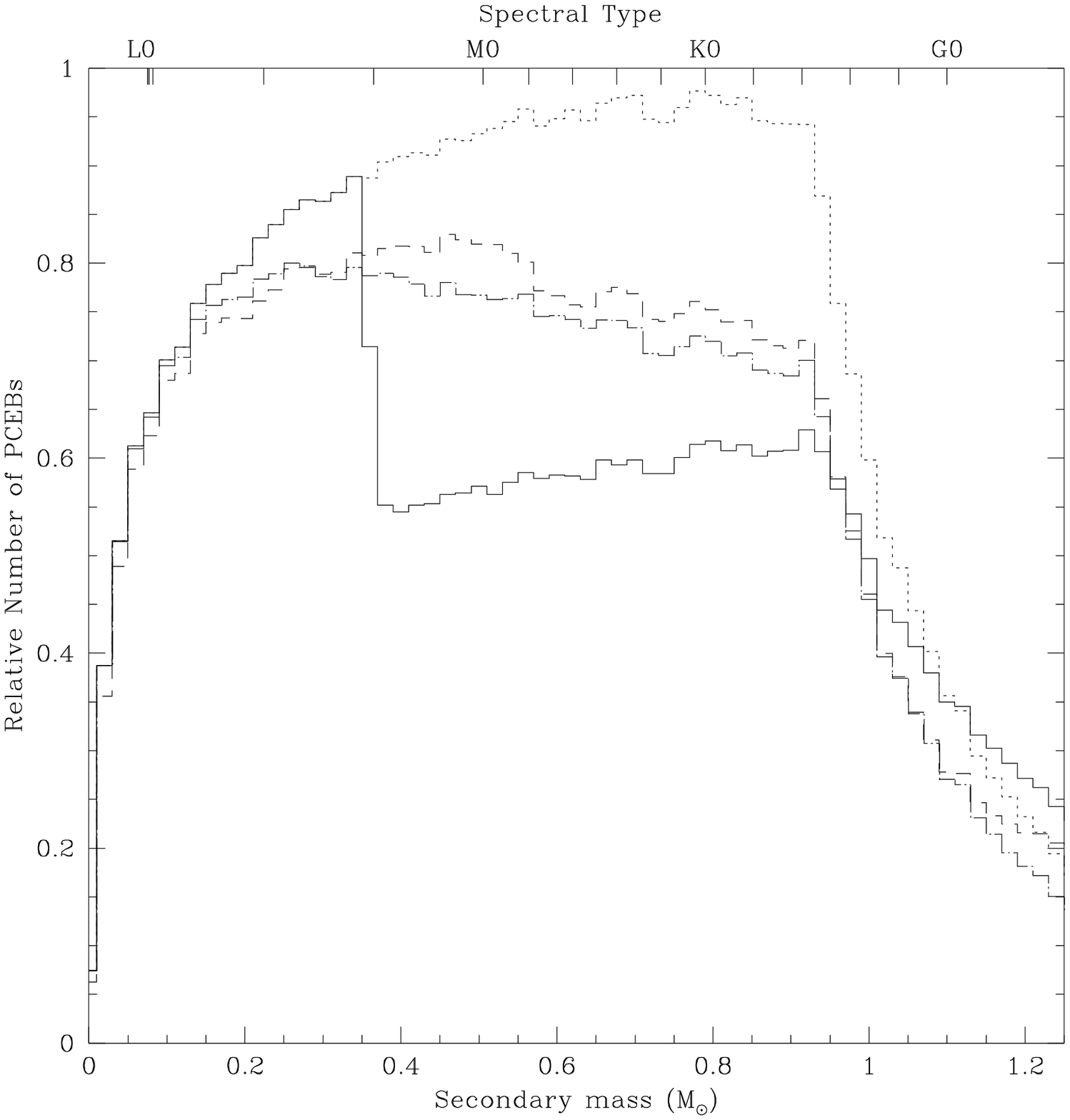}
\figcaption{Theoretical distributions of the secondary mass in present-day PCEBs for the four different assumed AML models described in the text:  GR only (dotted line), GR + RMB (dashed line), GR + IMB (dashed-dotted line), and GR + DMB (solid line).  The y-axis has been arbitrarily normalized to facilitate comparison.  Corresponding spectral types are shown along the top axis.  Spectral type-mass relationships from \citet{kir91} and \citet{kir94} were used for M dwarfs.} 
\end{figure}

\end{document}